\documentclass[a4paper]{article}

\usepackage[english]{babel}
\usepackage[utf8]{inputenc}
\usepackage{amsmath}
\usepackage{graphicx}
\usepackage[colorinlistoftodos]{todonotes}
\usepackage{authblk}
\usepackage{color}
\usepackage{ulem}
\RequirePackage{subfig}
\usepackage{hyperref}
\RequirePackage{amssymb}

\title{\textcolor{blue}{High Precision Calculations of the Higgs Boson Mass} }

\author[1]{ E. A. Reyes R.}
\affil[1]{ \small \textsl{ Universidad de Pamplona (UDP), Pamplona - Norte de Santander, Colombia }}
\author[2]{ A. R. Fazio}
\affil[2]{ \small \textsl{ Departamento de F\'{i}sica, Universidad Nacional de Colombia, Ciudad Universitaria, Bogot\'{a} D.C., Colombia}}
\date{}

\begin{document}
\maketitle

\begin{abstract}
In this paper we are going to review the status of the computations of the perturbative quantum corrections to the Higgs boson mass in the Standard Model and in its supersymmetric extensions. In particular, supersymmetric theories require a very accurate computation of the Higgs boson mass, which includes corrections up to even three-loop level, since their predictions are limited by theoretical uncertainties. A discussion about these uncertainties in the context of the minimal and next to minimal supersymmetric standard models is included.
\end{abstract}

\section{Introduction}

The Standard Model (SM) of the electro-weak forces unification does not explain important phenomena of fundamental interactions like the neutrino oscillation, the dark matter, the baryon asymmetry, the vacuum stability of the model, the electro-weak hierarchy in the mass scales, among others. The explanation of the above phenomena remains without a conclusive solution and requires new physics, so called beyond the Standard Model (BSM). Moreover recent experimental results are moving away from their SM expectations. At the beginning of this year, the Fermilab Muon g-2 collaboration showed an increasingly discrepancy between the measurement of the muon anomalous magnetic momentum and the corresponding SM theoretical prediction. The last report shows a $4.2 \sigma $ departure~\cite{B.Abi} from the reliable SM calculation, which are updated together with the inclusion of non-perturbative contributions in the so called white paper \cite{WP}.  Another recent result carried out by the LHCb collaboration, the so-called $R_K$ anomaly which is also related to the muon, reported the measurement of the decay rate of B-mesons to a kaon and a pair of muons compared to the decay into a kaon and electrons, providing evidence of the violation of lepton-flavor universality at $3 \sigma$ level~\cite{Roel}. \\ Assuming those discrepancies are a consequence of new physical degrees of freedom, it is important to study their relation with the other SM problems and the possible consequences of the physics beyond the Standard Model. The agenda of new colliders, such as the high-luminosity (HL-) LHC~\cite{HLLHC}, the International Linear Collider (ILC)~\cite{ILC1,ILC2}, the Future Circular Collider (FCC-ee)~\cite{FCC1,FCC2, FCC3}, or the Circular Electron Positron Collider (CEPC)~\cite{CEPC}, includes the design of experiments with the necessary capabilities to measure with higher accuracy the properties of the precision observables in order to look for new physics signals coming from the results that deviate from the SM predictions. There will be also the necessity to interpret the coming experimental results and study the implications for specific UV models. This includes the development and improvement of precision calculation techniques. New tools for the evaluation of amplitudes at three, four and even five -loops together with the development of techniques for the tensorial reduction of multi-leg strong (QCD) and electroweak (EW) mixed processes are needed to study the Higgs properties, the EW physics, the top, the QCD and the Heavy Flavor physics, the possible BSM production and the global interpretation of all the expected results in the future colliders . A survey of the present status of analytic calculation methods   of higher orders Feynman diagrams with zero or single mass scale is given in~\cite{Blum}. In particular, the recent discovery of a Higgs boson -like particle at the LHC~\cite{ATLASCMS} opened a window to the study of new phenomena. BSM models based on supersymmetry (SUSY) where the Higgs boson mass is a prediction of the theory, and their low-energy effective field theories (EFT), include additional neutral and charged particles in the Higgs sector with a mass above or below the energy scale where the LHC Higgs boson was found. The compatibility of this extended Higgs sector with the experimental results coming from the combined ATLAS/CMS measurements~\cite{ATLASCMS2} will require a SM-like Higgs at $125.09\pm 0.21\text{(stat.)}\pm0.11\text{(syst.)}~GeV$, but also an agreement of the properties of the new particles with the present experimental bounds coming from the searches of dark matter candidates, the muon $g-2$ anomaly, the LHCb results and the neutrino oscillations is mandatory. All of this phenomenological restrictions impose important constraints on the parameter space of the BSM models and can serve as a guide to look for the parameter region where the new particles might have appreciable interactions. In addition, it implies a very accurate computation of the Higgs boson mass since the central experimental value of $125.09~GeV$ is far to be accurate at tree-level in minimal supersymmetric models and besides the uncertainty associated to the theoretical predictions must reach the experimental accuracy at LHC and at future colliders, estimated to be of the order of $100~MeV$~\cite{Aguilar}. The high-precision calculations of the radiative corrections to the Higgs boson mass is therefore of major importance to supersymmetric phenomenology. \\ In this article we review the theoretical status of the Higgs boson mass in the SM and also in  the main BSM, supersymmetric extensions. We start by introducing in Section~\ref{HiggsSM} the Higgs sector of the SM at tree-level and then we review in Section~\ref{HiggsMSSM} the Higgs sector in the context of softly broken supersymmetry. We focus our attention on the minimal supersymmetric extension of the SM (MSSM) with real parameters. The expressions of the tree-level Higgs boson masses and their leading quantum corrections in the above mentioned models together with a discussion about the conditions to guarantee the stability of the Higgs potential are also included in Section~\ref{HiggsSM} and Section~\ref{HiggsMSSM}. In Section~\ref{HiggsNMSSM} a discussion about the theoretical uncertainties of the Higgs boson mass in supersymmetric models including some of the next to minimal extensions which enable to describe the recent experimental observations, as the one obtained in FERMILAB for the muon $g-2$ anomaly, is included. Finally, in Section~\ref{Conclusions} we give a research outlook and conclude.

\section{The Higgs Sector of the SM} \label{HiggsSM}
\begin{figure}
\centering
\includegraphics[width=16pc]{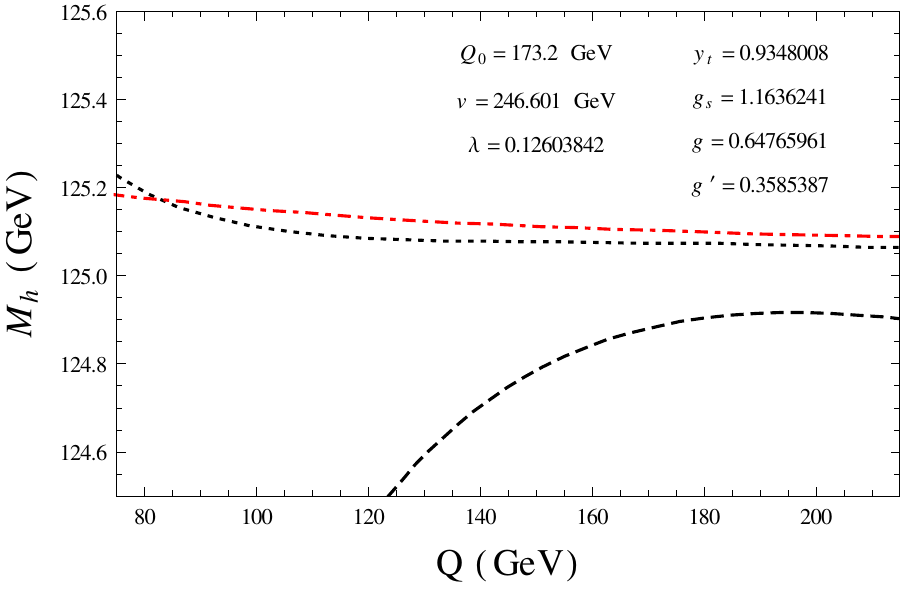}\hspace{2pc}%
\caption{\small{Renormalized Higgs boson mass in the tadpole-free $\overline{\text{MS}}$ scheme at full one-loop (dashed), full two-loop (dotted) and three-loop leading logarithms (dot-dashed red) level as a function of the renormalization scale $Q$.}}
\label{fig:PlotMhSM}
\end{figure}

In the SM, the Higgs sector contains just one scalar Higgs doublet
\begin{equation}
    H=\left(\begin{array}{c}
\phi^{+}\\
\phi^{0}
\end{array}\right),
\end{equation}
which is sufficient to give masses to the three generation of quarks, the three up-type leptons and the EW gauge bosons $Z^0$ and $W^\pm$. Neutrinos instead remain massless particles because of chiral symmetry. The method to generate the masses is the famous Higgs mechanism of electroweak symmetry breaking~\cite{ref-Higgs, ref-Englert} where the Higgs sector is introduced through the Lagrangian 
\begin{equation}
    \mathcal{L}_{H}=\left(D_{\mu}H\right)\left(D^{\mu}H\right)^{\dagger}-V(H). \label{LH}
\end{equation}
The first component of the Higgs doublet, $\phi^+ = \phi_1 + i\phi_2$, is a complex field that has electric charge $Q=1$ and weak isospin third component $T_3 = 1/2$, while the component $\phi^0 = \phi_3 + i\phi_4$ is a neutral complex field ($Q=0$) and has weak isospin equals to $T_3=-1/2$. As a consequence, both components have an $U(1)$ hypercharge $Y(H)=2(Q-T_3)=1$. The covariant derivative, $D_\mu$, is defined as
\begin{equation}
    D_{\mu}=\partial_{\mu}-i\frac{g}{2}\tau^{a}W_{\mu}^{a}-i\frac{g'}{2}B_{\mu},
\end{equation}
where $\tau^a$ are the Pauli's matrices, the parameters $g$ and $g'$ are the interaction couplings associated to the $SU(2)_L$ and $U(1)_Y$ gauge symmetries respectively. The electroweak gauge bosons are defined as 
\begin{eqnarray}
& & W_{\mu}^{\pm}=\frac{W_{\mu}^{1}\mp i\, W_{\mu}^{2}}{\sqrt{2}} \; ; \; \left(\begin{array}{c}
Z^0_{\mu}\\
A_{\mu}
\end{array}\right)=\left(\begin{array}{cc}
cos\theta_{W} & -sin\theta_{W}\\
sin\theta_{W} & cos\theta_{W}
\end{array}\right)\left(\begin{array}{c}
W_{\mu}^{3}\\
B_{\mu}
\end{array}\right), \label{eq:EWbosons}
\end{eqnarray}
where the EW mixing angle, $\theta_W$, is related to the coupling constants $g$ and $g'$ and the electric charge, $e$, by
\begin{eqnarray}
& & g=\frac{e}{sin\theta_W} \; ; \; g'=\frac{e}{cos\theta_W}. \label{eq:thetaW}
\end{eqnarray}
Additionally, the classical potential $V(H)$ is restricted by the gauge invariance and renormalizability to have the functional form:
\begin{equation}
    V(H)=\mu^{2}\left|H\right|^{2}+\lambda\left|H\right|^{4}.
\end{equation}
The Higgs quartic coupling $\lambda$ is a dimensionless positive number in order that the potential be bounded from below as $H$ goes to infinity \cite{Iliopoulos}, while the quadratic parameter $\mu^2$ can take either positive or negative values. In the case where $\mu^2$ is positive, the classical potential has a minimum at $H=0$, the $SU(2)$ symmetry of the Lagrangian is manifest and $\mu^2$ can be interpreted as the mass parameter of the field $H$. Note that the Lagrangian $\mathcal{L}_{H}$ is also invariant under $H\rightarrow -H$. On the other hand, if $\mu^2$ is negative the potential develops an infinite number of equivalent minima, determined by all the field configurations for which 
\begin{equation}
    \frac{\partial V(H)}{\partial|H|^{2}}=0\Rightarrow\left|H\right|^{2}=-\frac{\mu^{2}}{2\lambda} = \frac{1}{2}v^{2}.
\end{equation}
The new parameter $v$ is called the vacuum expectation value (vev) of $H$ and is defined up to a local gauge transformation under the gauge group $SU(2)$ of weak isospin. The election of one of those minima breaks spontaneously the gauge symmetry, the vacuum is not invariant under the symmetry group of the Lagrangean. In order to determine the particle spectrum, we have to choose one of the possible minimum configurations and expand the field $H$ around it. Due to the $SU(2)$ symmetry, we can define a new real scalar field $h(x)$ in the so called unitary gauge, such that $\phi_1 = \phi_2=0$, $\phi^0 = (1/\sqrt{2})(v+h)$ and therefore
\begin{equation}
    H(x)=\frac{1}{\sqrt{2}}\left(\begin{array}{c}
0\\
v+h(x)
\end{array}\right).
\end{equation}
The choice of $\phi^0$ as the component of the Higgs doublet that gets a vacuum expectation value ensures the conservation of the electric charge. Besides, since only one component gets a vev when one chooses a direction, we have three broken continuous symmetries of the original $SU(2)\times U(1)$ continuous symmetry. As a consequence of the Goldstone theorem, the resulting spectrum contains three massless, zero-spin Goldstone bosons, that will become the zero spin component of the weak gauge bosons $Z^0$ and $W^{\pm}$. Expanding the classical potential around $h=0$ we have:
\begin{equation}
    V(H)=\frac{1}{2}\left(2\lambda v^{2}\right)h^{2}+\lambda vh^{3}+\frac{1}{4}\lambda h^{4}.
\end{equation}
It is possible to associate a physical particle to the quantum field $h(x)$ with a tree-level mass proportional to the self-interaction coupling $\lambda$ and to the vev of $H$: 
\begin{equation}
    M_h = \sqrt{2\lambda} v.
\end{equation}
This neutral zero-spin particle is well known as the SM Higgs boson and so far, its properties and all the SM predictions are in a great agreement with the Higgs boson properties measured at LHC by the ATLAS and CMS experiments. It is worth to mention that the mass of the Higgs boson in the SM is not a prediction of the theory, but a free input parameter whose numerical value must be specified in the theoretical computations of the other $M_h$-dependent SM observables. \\ From the kinetic term of the Lagrangian in eq. (\ref{LH}) and after expanding $H$ around its vev, we have:
\begin{equation*}
    \left(D_{\mu}H\right)\left(D^{\mu}H\right)^{\dagger}=\frac{1}{2}\partial_{\mu}h\partial^{\mu}h+\left[\frac{1}{4}g^{2}v^{2}W_{\mu}^{+}W^{\mu-}+\frac{1}{8}(g^{2}+g'^{2})v^{2}Z_{\mu}Z^{\mu}\right]\left(1+\frac{h}{v}\right)^{2}.
\end{equation*}
Whereas the Lagrangian develops additional mass terms for the EW gauge bosons, the photon remains massless and therefore
\begin{equation}
    M_W = \frac{1}{2} vg;\quad  M_Z = \frac{1}{2}v\sqrt{g^2+g'^2}; \quad M_{\gamma} = 0. 
\end{equation}
From the relation of the gauge boson mass $M_W$ with the Fermi coupling constant obtained from the muon decay process,   
\begin{equation}
    \frac{G_F}{\sqrt{2}} = \frac{g^2}{8M_W^2}(1 + \Delta r),
\end{equation}
where the Fermi constant $G_F\approx 1.17\times 10^{-5}~GeV^{-2}$ and $\Delta r$ contains the radiative corrections to the relation, we can estimate the tree-level value for the vev of $H$, 
\begin{equation}
    v = \sqrt{\frac{1}{\sqrt{2} G_F}} \approx 246 ~ GeV. 
\end{equation}
Note that the mass generation for the EW bosons through the Higgs mechanism does not spoil the gauge invariance of the Lagrangian and consequently the theory is renormalizable, contrarily to what happens in the Proca theory, where the vector bosons mass is a given parameter of the bare Lagrangian~\cite{Becchi}. On the other side, quark masses require the existence of left-handed $SU(2)$ doublets and right-handed $SU(2)$ singlets in such a way that
\begin{equation}
    \mathcal{L}_q = y_d \overline{Q}_L H d_R + y_u \overline{Q}_L H_c u_R + h.c.
\end{equation}
where
\begin{equation}
    \overline{Q}_{L}=\left(\begin{array}{c}
u\\
d
\end{array}\right)_{L};\qquad H_{c}=-\frac{1}{\sqrt{2}}\left(\begin{array}{c}
v+h(x)\\
0
\end{array}\right),
\end{equation}
for quarks type up, $u$, and type down, $d$. The Yukawa parameters $y_u$ and $y_d$ are arbitrary, thus, the Higgs mechanism produces expressions for quark masses in terms of the Higgs vev, 
\begin{equation}
    m_{u,d} = \frac{y_{u,d}}{\sqrt{2}}v. 
\end{equation}
Similarly, masses of the leptons are generated from the interaction Lagrangian of the leptons with the Higgs field,
\begin{equation}
    \mathcal{L}_{L}=y_{e}\left(\overline{L}He_{R}+H^{\dagger}\overline{e}_{R}L\right),
\end{equation}
where $y_e$ is the Yukawa coupling for the $e$-type lepton ($e=e^-,\mu,\tau$) $L$ is an $SU(2)$ doublet with components that contain a left-handed lepton, $e_L$, and a left-handed neutrino, $\nu_{e_L}$,
\begin{equation}
    L=\left(\begin{array}{c}
\nu_{e}\\
e^{-}
\end{array}\right)_{L},
\end{equation}
and $e_R$ are right-handed leptons which are $SU(2)$ singlets. The mass term for the lepton $e$ is of the form:
\begin{equation}
    m_e = \frac{y_e}{\sqrt{2}} v.
\end{equation}
Due to the absence of a right-handed neutrino $\nu_{e_R}$, there is no mass term for $\nu_e$.  Experimentally, only left-handed neutrinos have been observed with masses that are very small but non-zero. The masses of the quarks and up-type leptons are not predicted in the model and therefore, just like the Higgs mass, they are input (running) parameters that must be measured at the experiments. The experimental measured values are related to the pole masses by the spectral representation of the full propagator.
For the Higgs boson, the renormalized mass in the $\overline{\text{MS}}$ scheme has been computed to two-loop order including the mixed QCD/EW contributions to $M_h$ with the explicit inclusion of the tadpoles in the Higgs self-energy diagrams~\cite{Kniehl1}, also including the O($\alpha\alpha_s$) corrections to the relation between the $\overline{\text{MS}}$ and pole masses of the Higgs boson. The radiative corrections in the gauge-less limit approximation at two-loop order is presented in~\cite{Degrassi}, while the full two-loop correction in a hybrid $\overline{\text{MS}}$/on-shell scheme can be found in~\cite{Buttazzo}. In addition, the Higgs boson mass in the tadpole-free pure $\overline{\text{MS}}$ scheme including all the one-loop and two-loop contributions can be consulted in \cite{Martin1}, where also the leading three-loop contributions estimated using the 1PI effective potential approach in the limit where $g^2,g'^2,\lambda \ll g_s^2,y_t^2$ is included. The latter corrections together with the relations between the $\overline{\text{MS}}$ parameters and the on-shell masses of the SM are implemented in the public code \textit{SMDR}~\cite{Martin2}. With the help of \textit{SMDR} the state of art of the Higgs boson mass in the SM can be easily visualized. Figure~\ref{fig:PlotMhSM} shows the Higgs boson pole mass in the $\overline{\text{MS}}$ scheme as a function of the renormalization scale, $Q$, obtained from the \textit{SMDR} code. The plot includes the full one-loop contribution (dashed black curve), full two-loop contribution (dotted black line) and the leading three-loop corrections (dot-dashed red line) at order O($g_s^4y_t^2$)+O($g_s^2y_t^4$)+O($y_t^6$), where $g_s$ is the strong coupling constant.   
The curves were drawn using the default input values in \textit{SMDR} of the $\overline{\text{MS}}$ parameters at the initial scale $Q_0=173.2~GeV$. The running parameters of the model at $Q$ are determined by the renormalization group evolution from the subtraction point defined at $Q_0$. Note that the renormalization scale dependence makes the difference between the three loop and tree level result of the order of several tenths of MeV in Figure~\ref{fig:PlotMhSM}  when $Q$ is close to $160~GeV$. Nevertheless, as was argued in~\cite{Martin1}, this error can increase if we are far away of this scale. We must take into account that the three-loop correction to $M_h$ included in \textit{SMDR} is in the 1PI effective potential approximation where the external momentum of the Higgs self-energies vanishes. The full determination of the complex pole squared mass requires, however, the inclusion of the external momentum dependence of the Higgs self energy. A complete Feynman diagrammatic computation of the three-loop Higgs self-energies is expected to make flat the $Q$ dependence. Besides, even though the effective potential approximation might be expected to be the bulk of the three-loop correction, the complete two loop order with leading three-loop corrections of the pole mass  in ~\cite{Martin1} may receive significant corrections because the ratio $M_h/M_t\approx 0.6$ is not a really small expansion parameter. An additional important issue has to do with the stability of the SM vacuum. The LHC value of $M_h$ implies a near critical metastable vacuum~\cite{Degrassi, Buttazzo}, the metastability is the preferred option with $99.3$\% of probability, which means that the SM is at the border between stability and metastability regions and could be sitting on the stability region, i.e. it could reach and even cross the stability line, when its experimental and theoretical uncertainties have been taken into account. Including the two-loop threshold corrections of all SM parameters, the estimated overall theory error on $M_h$ is about $1.0~GeV$. Although the dominant parametric uncertainties come from the top-quark mass and the QCD coupling, the limitations of the knowledge about the state of the SM vacuum should rely mainly on experimental errors. An unambiguously theoretical computation of $M_h$ should, to the possible extent, contains a theoretical error which reach at least the current experimental uncertainty. It is therefore important to perform the computation of the Higgs boson mass and its relation with the SM parameters as accurately as possible. Additionally, the information from the SM masses can be inverted to obtain the evolution of the interaction couplings constants. This can be done with the code \textit{SMDR} as well, but there is also an alternative code named \textit{mr}~\cite{mr} which is based on the calculations reported in~\cite{Kniehl2} that contains the relationships between the SM parameters of the on-shell scheme and their counterparts in the $\overline{\text{MS}}$ scheme at full two-loop order. \\ It is worth mentioning that in \textit{mr} the definition and renormalization of the vev is different from the used in \textit{SMDR}. In \textit{mr}, the renormalized vacuum, $v_{tree}$, is defined as the minimum of the tree-level Higgs potential. The advantage of a scheme with this vacuum definition is that all the physical observables depending on it are gauge invariant  due to the explicit insertion of the tadpole diagrams. Their insertion is done not only in the diagrams with counterterms but also in the 1PI Feynman diagrams as required to have the gauge independence of renormalized Green functions according to the gauge Ward identities. However, there is a disadvantage, the inclusion of tadpole diagrams involving the Higgs field include negative powers of the Higgs mass leading to a parametrically slower convergence. The leading small-$M_h$ behaviour is dominated by the one-loop correction, the threshold corrections to $M_h$ and $\lambda$ contain terms that diverge as $1/M_h^2$ for $M_h\rightarrow 0$~\cite{Kniehl2, Sirlin}. In \textit{SMDR} instead, the vev is defined to be the minimum of the full effective potential calculated in the Landau gauge. By this definition, the sum of all Higgs tadpole graphs, including the tree-level Higgs tadpole, vanishes identically, and therefore negative powers of $\lambda$ and huge EW corrections are absent in the perturbative expansions of the pole masses and their relations with the $\overline{\text{MS}}$ parameters. This vev is in some sense a more faithful description of the true vacuum state. However, there is also a disadvantage. If the scheme is defined in terms of the self-energy diagrams without including the tadpole contribution, it gives rise to radiative corrections that are dependent of the gauge-fixing condition~\cite{Fleischer}. As a consequence, in this framework the $\overline{\text{MS}}$ masses are gauge-dependent quantities. Nevertheless, a $\overline{\text{MS}}$ mass is not a physical quantity nor a parameter of the bare Lagrangian, moreover the physical observable are independent on the running of the renormalized parameters (up to the error of the truncation of the perturbative series) and therefore the requirement of gauge-invariance for the running of the $\overline{\text{MS}}$ mass  is not mandatory. The precise relationship between the threshold corrections that relate the $\overline{\text{MS}}$ masses to the parameters in the pole scheme with the three level vacuum $v_{tree}$ and the alternative computation implemented in \textit{SMDR} deserve more attention for future studies. 

\section{Higgs Sector in Supersymmetry} \label{HiggsMSSM}
We start by reviewing the Higgs sector of the minimal supersymmetric extension of the SM (MSSM) at tree-level. Supersymmetry is the only fermionic generalization of the Poincaré symmetry of the space-time which transforms fermions into bosons and vice versa. The existence of such a non-trivial extension of the Poincaré symmetry is highly constrained by theoretical principles \cite{Poincare1,Poincare2}. The MSSM contains the fields of the two-Higgs-doublet extension of the SM (2HDM) and their corresponding superpartners. The field content of the MSSM is summarized in Table~\ref{tb:SUSYspectrum} \cite{Langacker}. 
\begin{table}
\centering
{\footnotesize{}
\begin{tabular}{|c|c|c|c|c|}
\hline 
Super-Multiplet & Super-field & Bosons & Fermions & \footnotesize{$SU(3)\times SU(2) \times U(1)$} \tabularnewline
\hline 
\hline 
gauge/ino & $\begin{array}{c}
\widehat{V}_{8}\\
\widehat{V}\\
\widehat{V}'
\end{array}$ & $\begin{array}{c}
g\\
W^{\pm},\;W^{0}\\
B
\end{array}$ & $\begin{array}{c}
\widetilde{g}\\
\widetilde{W}^{\pm},\;\widetilde{W}^{0}\\
\widetilde{B}
\end{array}$ & $\begin{array}{c}
8\oplus 1\oplus 0\\
1\oplus 3 \oplus 0\\
1 \oplus 1 \oplus 0
\end{array}$ \tabularnewline
\hline 
s/lepton & $\begin{array}{c}
\widehat{L}\\
\widehat{E}^{c}
\end{array}$ & $\begin{array}{c}
\left(\widetilde{v}_{L},\;\widetilde{e}_{L}^{-}\right)\\
\widetilde{e}_{R}^{+}
\end{array}$ & $\begin{array}{c}
\left(v,\;e^{-}\right)_{L}\\
e_{L}^{c}
\end{array}$ & $\begin{array}{c}
1 \oplus 2 \oplus ( -1)\\
1 \oplus 1 \oplus 2
\end{array}$ \tabularnewline
\hline 
s/quark & $\begin{array}{c}
\widehat{Q}\\
\widehat{U}^{c}\\
\widehat{D}^{c}
\end{array}$ & $\begin{array}{c}
\left(\widetilde{u}_{L},\;\widetilde{d}_{L}\right)\\
\widetilde{u}_{R}^{\ast}\\
\widetilde{d}_{R}^{\ast}
\end{array}$ & $\begin{array}{c}
\left(u,\;d\right)_{L}\\
u_{L}^{c}\\
d_{L}^{c}
\end{array}$ & $\begin{array}{c}
3 \oplus 2 \oplus 1/3\\
\bar{3} \oplus 1 \oplus (-4/3)\\
\bar{3} \oplus 1 \oplus 2/3
\end{array}$ \tabularnewline
\hline 
Higgs/ino & $\begin{array}{c}
\widehat{H}_{1}\\
\widehat{H}_{2}
\end{array}$ & $\begin{array}{c}
\left(H_{1}^{0},\;H_{1}^{-}\right)\\
\left(H_{2}^{+},\;H_{2}^{0}\right)
\end{array}$ & $\begin{array}{c}
\left(\widetilde{H}_{1}^{0},\;\widetilde{H}_{1}^{-}\right)\\
\left(\widetilde{H}_{2}^{+},\;\widetilde{H}_{2}^{0}\right)
\end{array}$ & $\begin{array}{c}
1 \oplus 2 \oplus (-1)\\
1 \oplus 2 \oplus 1
\end{array}$ \tabularnewline
\hline 
\end{tabular} }{\footnotesize\par}
\caption{\small{MSSM spectrum fields and their $SU(3)\times SU(2)\times U(1)$ quantum numbers. In this table only one generation of leptons and quarks is exhibited. For each lepton and quark super-multiplet there is associated a corresponding multiplet of scalar superpartners, while for the Higgs super-multiplet an anti-particle multiplet of charge-conjugated fermions is associated.}}
\label{tb:SUSYspectrum}
\end{table} 
Specifically, the spectrum contains three gauge supermultiplets which consist of the gluons with their corresponding gluino fermionic superpartners and the EW gauge bosons with their gaugino superpartners. The matter supermultiplets consist of three generations of left-handed quarks and their scalar superpartners, the squarks, of three generations of left-handed leptons and their associated sleptons, and the corresponding antiparticles of all the fermions. The Higgs supermultiplets consists of two complex Higgs doublets, their higgsino fermionic superpartners and the corresponding antiparticles. In order to guarantee the cancellation of anomalies coming from the introduction of the higgsino superpartners and preserve supersymmetry and gauge invariance, the Higgs sector of the MSSM Lagrangian \cite{Fayet,Nilles,HabernKane,Drees,Kuroda,Terning,Baer,Aitchison} requires the definition of at least two doublets
\begin{eqnarray}
& & H_{1}=\left(\begin{array}{c}
H_{1}^{0}+\frac{1}{\sqrt{2}}v_{1}\\
H_{1}^{-}
\end{array}\right) \;\; \mathrm{and} \;\; H_{2}=e^{i\psi_2}\left(\begin{array}{c}
H_{2}^{+}\\
H_{2}^{0}+\frac{1}{\sqrt{2}}v_{2}
\end{array}\right), \label{eq:Doublets}
\end{eqnarray}
with associated hypercharges $Y_1 = -1$ and $Y_2 = +1$, respectively. The vacuum expectation values $v_1$ and $v_2$ have a phase difference which is parametrized through the angle $\psi_2$. In the real version of the MSSM (rMSSM) this angle is removed through a hypercharge rotation. We assume in the following that $\psi_2=0$. The second Higgs doublet is necessary, moreover, to generate mass for both up- and down-type quarks as well as charged leptons. The complex fields $H_j^{0}$ have neutral components which are vevless scalar fields, $H_j^0 = \phi_j^0 + i\chi_j^0$, coming from the expansion around the minima of the Higgs effective potential, $v_{1,2}$. The components $H_{1,2}^{\pm}$ are charged scalar fields with vanishing vacuum expectation values (vevs) to ensure that the $U(1)_{em}$ symmetry will not be spontaneously broken. The Higgs sector comes from the bare Lagrangian
\begin{eqnarray}
& & \mathcal{L}_{V-H}=\mathcal{D}_{\sigma}H_{1}^{\dagger}\mathcal{D}^{\sigma}H_{1}+\mathcal{D}_{\sigma}H_{2}^{\dagger}\mathcal{D}^{\sigma}H_{2}-V(H_{1},H_{2}).\label{eq:LagHiggs}
\end{eqnarray}  
The kinetic term contains the covariant derivative 
\begin{eqnarray}
& & \mathcal{D}_{\sigma} = \partial_{\sigma} + ig\frac{\tau^a}{2}W_{\sigma}^{a} + ig'\frac{Y}{2}B_{\mu} + ig_s\frac{\lambda^{\alpha}}{2}G_{\sigma}^{\alpha},
\end{eqnarray}
where $g'$, $g$ and $g_s$ are the U$(1)_Y$, SU$(2)_L$ and SU$(3)$ gauge couplings respectively. The SU$(2)$ generators Pauli matrices are represented by $\tau^a$ while $\lambda^{\alpha}$ stands for the SU$(3)$ Gell-Mann matrices. The classical Higgs potential, $V(H_{1},H_{2})$, is given by
{\footnotesize{}
\begin{eqnarray}
\quad ~~ V(H_{1},H_{2})=\left(\left|\mu\right|^2 + m_{H_{1}}^{2}\right)\left|H_{1}\right|^{2}+\left(\left|\mu\right|^2 + m_{H_{2}}^{2}\right)\left|H_{2}\right|^{2}+b\left(\epsilon_{ab}H_{1}^{a}H_{2}^{b}+h.c.\right)\nonumber \\ 
 ~~+~~\frac{1}{2}g^{2}\left|H_{1}^{\dagger}H_{2}\right|^{2}+\frac{1}{8}\left(g^{2}+g'^{2}\right)\left(\left|H_{2}\right|^{2}-\left|H_{1}\right|^{2}\right)^{2}.\label{eq:Exp-H-Pot}
\end{eqnarray}
}{\footnotesize\par}
The quadratic terms proportional to the higgsino mass parameter $\left|\mu\right|^2$ in eq.~(\ref{eq:Exp-H-Pot}) come from the $F$-contribution to the SUSY Lagrangian, while the terms with the EW gauge couplings ($g$, $g'$) come from the $D$-contribution; thus, from the $D$-terms, one obtains the four scalar vertices, which include the quartic Higgs self-interaction terms in the Higgs potential. In contrast to the SM, the Higgs quartic self-coupling is not a free parameter but is determined by the coupling $(g^2 + g'^2)/8$. From the $F$-terms, one obtains also another four scalar vertices, but they do not contribute to the Higgs potential since contain always at least two sfermions. The $F$ and $D$ fields are auxiliary fields that do not propagate in space-time and can be eliminated by applying the equations of motion. As supersymmetric particles have not been observed at the electroweak scale, supersymmetry is manifestly not an exact symmetry of the nature and therefore must be broken. Several SUSY breaking mechanisms have been considered \cite{Witten, Dimopoulos, Girardello} but in fact there is no consensus on how to break SUSY. In phenomenological application, one must introduce new terms that explicitly breaks supersymmetry and represent the low-energy effects of the unknown breaking mechanism. In eq. (\ref{eq:Exp-H-Pot}) the parameters $m_{H_{1}}^{2}$, $m_{H_{2}}^{2}$ and $b$ parameterize the soft SUSY breaking. The SUSY-invariant Lagrangian cannot accommodate electroweak symmetry breaking since the terms proportional to $|\mu|^2$ are positive. Thus, the SUSY-violating parameters $m_{H_{1}}^{2}$ and $m_{H_{2}}^{2}$, which can of course have either sign, are needed in order to break the EW symmetry. The $b$-term is the only that depends on the phases of the fields. The parameter $b$ is real and positive and the fields $H_1^0$ and $H_2^{0}$ have equal and opposite phases which can be reduced both to zero through a U$(1)_Y$ gauge transformation since these fields have equal and opposite hypercharges. The vevs, $v_{1,2}$, as well as the couplings are therefore all real, which means that CP invariance is not spontaneously broken by the classical Higgs potential of the rMSSM. \\ The softness of the SUSY breaking terms means that they must be super-renormalizable, with mass dimension less than four, and therefore their couplings must have positive mass dimension. The reason is that a soft breaking term will not introduce additional divergences into the dimensionless coupling constants which guarantee the stability of the mass hierarchy, the cancellation of quadratically divergent corrections to scalar masses are maintained to all perturbative orders. Besides, it is important to emphasize that the mass terms which break SUSY and therefore the masses of the undiscovered SUSY particles do respect the SM gauge symmetries. The masses of the known SM particles all arise from the spontaneous breaking of the EW symmetry. \\
From eq. (\ref{eq:Exp-H-Pot}) we can derive the linear part of the Higgs potential in the basis ($\phi_j^0, \chi_j^0, H_{j}^{\pm}$) where $j=1,2$. For the fields $\chi_j^0$ and $H_{j}^{\pm}$ there are no contributions since the rMSSM Higgs potential is invariant under CP-tranformation. Thus, the linear term of the effective potential is $T_1\phi^{0}_1 + T_2\phi^{0}_2$, where the coefficients $T_j$, better-known as Higgs tadpoles, have the expressions
\begin{eqnarray}
& & \frac{T_1}{\sqrt{2}v_1} = \left(\left|\mu\right|^2+m_{H_{1}}^{2}\right)-b\frac{v_{2}}{v_1}+\frac{1}{8}\left(g^{2}+g'^{2}\right)\left[v_{1}^{2}-v_{2}^{2}\right], \nonumber \\
& & \frac{T_2}{\sqrt{2}v_2} = \left(\left|\mu\right|^2+m_{H_{2}}^{2}\right)-b\frac{v_{1}}{v_2}+\frac{1}{8}\left(g^{2}+g'^{2}\right)\left[v_{2}^{2}-v_{1}^{2}\right]. \label{eq:Tadpoles}
\end{eqnarray}
As the vevs $v_1$ and $v_2$ minimize the Higgs potential, the following stationary conditions are satisfied:
\begin{eqnarray}
& & \left. \frac{\partial V}{\partial |H_j^0|} \right|_{\left\langle H_j^0 \right\rangle = 0 ;~ \left\langle H_j^{\pm} \right\rangle =0} = T_j = 0 \; ; \; j = 1,2 ~. \label{eq:mincond}
\end{eqnarray}
From eq. (\ref{eq:mincond}) one can identify the conditions required for the stable minimum of $V$. First note that along the direction $|H_1^0| = |H_2^0|$ the potential will be unbounded from below and therefore it does not have a minimum unless
\begin{eqnarray}
& & 2\left|\mu\right|^2 + m_{H_{1}}^{2} + m_{H_{2}}^{2} > 2b. \label{eq:stacon1}
\end{eqnarray}
For $\left(\left|\mu\right|^2+m_{H_{1}}^{2}\right)$ and $\left(\left|\mu\right|^2+m_{H_{2}}^{2}\right)$ positive, the origin is not a minimum but a saddle point, and the minimum occur at non-zero vevs of $H_j^0$, when 
\begin{eqnarray}
& & \left(\left|\mu\right|^2+m_{H_{1}}^{2}\right)\left(\left|\mu\right|^2+m_{H_{2}}^{2}\right) < b^2. \label{eq:stacon2}
\end{eqnarray}
The rMSSM Higgs potential develops a stable minimum if the conditions of the eq.~(\ref{eq:stacon1}) and eq.~(\ref{eq:stacon2}) are met. \\ Turning to the bilinear part of the kinetic terms in the Lagrangian $\mathcal{L}_{V-H}$ (eq. \ref{eq:LagHiggs}) the masses of the EW gauge bosons amount to
\begin{eqnarray}
& & M_W^2 = \frac{1}{4}g^2\left( v_1^2 + v_2^2 \right)~; \;\; M_{Z}^2 = \frac{1}{4}(g^2 + g'^2)\left( v_1^2 + v_2^2 \right)~; \;\; M_{\gamma}^2 = 0. 
\label{eq:EWmasses}
\end{eqnarray}
The $Z$ boson mass determines the tree-level relation $\sqrt{v_1^2 + v_2^2} \approx 174~GeV$. The mass matrices of the rMSSM Higgs bosons can be identified from the bilinear part of the classical potential,
{\footnotesize{}
\begin{eqnarray*}
\left(\begin{array}{cc}
\phi_{1}^{0} & \phi_{2}^{0}\end{array}\right)M^{\phi^{0}}\left(\begin{array}{c}
\phi_{1}^{0}\\
\phi_{2}^{0}
\end{array}\right)+\left(\begin{array}{cc}
\chi_{1}^{0} & \chi_{2}^{0}\end{array}\right)M^{\chi^{0}}\left(\begin{array}{c}
\chi_{1}^{0}\\
\chi_{2}^{0}
\end{array}\right)+\left(\begin{array}{cc}
H_{1}^{+} & H_{2}^{+}\end{array}\right)M^{H^{\pm}}\left(\begin{array}{c}
H_{1}^{-}\\
H_{2}^{-}
\end{array}\right). \label{eq:Hmass}
\end{eqnarray*}
}{\footnotesize\par}
To derive the $M^{\phi^{0}}$-matrix the relations from eqs. (\ref{eq:Tadpoles}) and (\ref{eq:mincond}) and the definitions
\begin{eqnarray}
& & M_{A}^{2}=b\left(cot\beta+tan\beta\right), \;\; tan\beta=\dfrac{v_{2}}{v_{1}}~; \;\; 0\leq\beta\leq\frac{\pi}{2}, \label{eq:Param}
\end{eqnarray}
are required. The tree-level mass matrix of the neutral $\phi^0_{1,2}$-bosons reads 
\begin{eqnarray}
& & M^{2}_{\phi^{0}} = \left(\begin{array}{cc}
M_{Z}^{2}c_{\beta}^{2}+M_{A}^{2}s_{\beta}^{2}+\dfrac{T_{1}}{\sqrt{2}v_{1}} & -\left(M_{A}^{2}+M_{Z}^{2}\right)s_{\beta}c_{\beta}\\
-\left(M_{A}^{2}+M_{Z}^{2}\right)s_{\beta}c_{\beta} & M_{Z}^{2}s_{\beta}^{2}+M_{A}^{2}c_{\beta}^{2}+\dfrac{T_{2}}{\sqrt{2}v_{2}}
\end{array}\right). \label{eq:phimass}
\end{eqnarray}
We have used the short notation $s_\beta=sin(\beta)$ and $c_\beta=cos(\beta)$ and we have written explicitly the contributions of the Higgs tadpoles, which vanish at tree-level according to eq. (\ref{eq:mincond}), because they will receive non-zero loop contributions when renormalization of the Higgs masses will be considered. Because both $v_1$ and $v_2$ are real and positive, the upper and lower bound on the angle $\beta$ lies on the interval shown in eq. (\ref{eq:Param}). By other side, the tree-level mass matrices of the $\chi_{1,2}^0$ and $H_{1,2}^{\pm}$ components are given by 
\begin{eqnarray}
& & M^{2}_{\chi^{0}}=\left(\begin{array}{cc}
M_{A}^{2}s_{\beta}^{2}+\dfrac{T_{1}}{\sqrt{2}v_{1}} & -M_{A}^{2}s_{\beta}c_{\beta}\\
-M_{A}^{2}s_{\beta}c_{\beta} & M_{A}^{2}c_{\beta}^{2}+\dfrac{T_{2}}{\sqrt{2}v_{2}}
\end{array}\right) \label{eq:chimass}
\end{eqnarray}
and
\begin{eqnarray}
& & M^2_{H^{\pm}}=\left(\begin{array}{cc}
M_{A}^{2}s_{\beta}^{2}+\dfrac{T_{1}}{\sqrt{2}v_{1}}+\frac{1}{2}g^{2}v_{2}^{2} & M_{A}^{2}s_{\beta}c_{\beta}+\frac{1}{2}g^{2}v_{1}v_{2}\\
M_{A}^{2}s_{\beta}c_{\beta}+\frac{1}{2}g^{2}v_{1}v_{2} & M_{A}^{2}c_{\beta}^{2}+\dfrac{T_{2}}{\sqrt{2}v_{2}}+\frac{1}{2}g^{2}v_{1}^{2}
\end{array}\right). \label{chargedmass}
\end{eqnarray}
The potential can be brought into the physical basis, where the quadratic terms in the components of $H_j$ are diagonalized, through the rotations 
\begin{eqnarray*}
\left(\begin{array}{c}
\phi_{1}^{0}\\
\phi_{2}^{0}
\end{array}\right)=D^{\dagger}_\alpha\left(\begin{array}{c}
H\\
h
\end{array}\right), & 
\left(\begin{array}{c}
\chi_{1}^{0}\\
\chi_{2}^{0}
\end{array}\right)=D^{\dagger}_\beta\left(\begin{array}{c}
G^{0}\\
A
\end{array}\right), & 
\left(\begin{array}{c}
H_{1}^{\pm}\\
H_{2}^{\pm}
\end{array}\right)=D^{\dagger}_\beta\left(\begin{array}{c}
G^{\pm}\\
H^{\pm}
\end{array}\right), \label{eq:Pi-A}
\end{eqnarray*}
via the orthogonal transformation
\begin{eqnarray}
& & D_\theta=\left(\begin{array}{cc}
c_{\theta} & s_{\theta}\\
-s_{\theta} & c_{\theta}
\end{array}\right). \label{eq:D(theta)}
\end{eqnarray}
In this basis the Higgs sector has five physical Higgs bosons, three of them are neutral: the lightest ($h$) and heavy ($H$) CP-even Higgs bosons and the CP-odd Higgs boson ($A$). The other two, $H^{\pm}$, are charged and vevless. There are also three unphysical massless Goldstone bosons, $G^{0}$ and $G^{\pm}$, which are absorbed by the EW gauge fields to build up their longitudinal components just as in the SM. The angle $\beta$ is linked to the vevs through eq. (\ref{eq:Param}) while $\alpha$ can be determined from the rotation of eq. (\ref{eq:phimass}) in terms of the MSSM parameters,
\begin{eqnarray}
& & tan(2\alpha)=tan(2\beta) \frac{ M_{A}^{2} + M_{Z}^{2}}{M_{A}^{2}-M_{Z}^{2}} \quad ; \quad  -\frac{\pi}{2} < \alpha < 0 ~ . \label{eq:alpha}  
\end{eqnarray}
Thus, after diagonalization, besides of the EW boson masses of eq. (\ref{eq:EWmasses}), the rMSSM Higgs sector is parametrized in terms of two additional parameters: $tan\beta$ and the mass of the CP-odd Higgs boson $m_{A}$. The masses of the charged Higgs bosons, $m_{\pm}$, are linearly dependent on $m_A$, they are usually used in the complex version of the MSSM. At tree-level we have:
\begin{eqnarray}
& & m_A^2 = M_{A}^{2}+\frac{T_{1}}{\sqrt{2}v_{1}}s_{\beta}^{2}+\frac{T_{2}}{\sqrt{2}v_{2}}c_{\beta}^{2} \quad ; \quad 
m_{\pm}^2 = m_A^2 + M_W^2. \label{eq:mAandmpm}
\end{eqnarray}
The tree-level masses of the CP-even Higgs boson particles, $h$ and $H$, follow as predictions
\begin{eqnarray}
& & m_{h,H}^{2}=\frac{1}{2}\left[m_{A}^{2}+M_{Z}^{2}\mp\sqrt{\left(m_{A}^{2}+M_{Z}^{2}\right)^{2}-4m_{A}^{2}M_{Z}^{2}cos^{2}\left(2\beta\right)}\right]. \label{eq:hHmasses}
\end{eqnarray}
In most of the relevant phenomenology benchmark scenarios for MSSM Higgs boson searches, the LHC Higgs boson corresponds to the lightest CP-even Higgs boson with a mass $m_h$ which is not a free input parameter but it is predicted in the MSSM. From the mass formulas (\ref{eq:mAandmpm}) and (\ref{eq:hHmasses}) the next important inequalities can be derived: 
\begin{eqnarray}
& & m_{h}\leq M_{Z}~; \quad m_{A}\leq m_{H}~; \quad M_{W}\leq m_{\pm}.
\end{eqnarray}
As a consequence, the lightest Higgs boson mass is predicted to be bounded from above by the $Z$-boson mass, $m_h\leq 91.2~GeV$, modulo radiative corrections. This bound follows from the fact that the quartic coupling of the Higgs boson fields is determined by the size of the EW gauge couplings. The tree-level bound on $m_h$ has already been excluded by the experimental value found at the LHC. However, this tree-level prediction is strongly modified by higher-order quantum corrections making the MSSM compatible with the measured Higgs mass of $125~GeV$ and consistent with the similarities of the measured Higgs couplings to those in the SM~\cite{ATLASCMS}. 
\begin{figure}
\centering
\includegraphics[width=16pc]{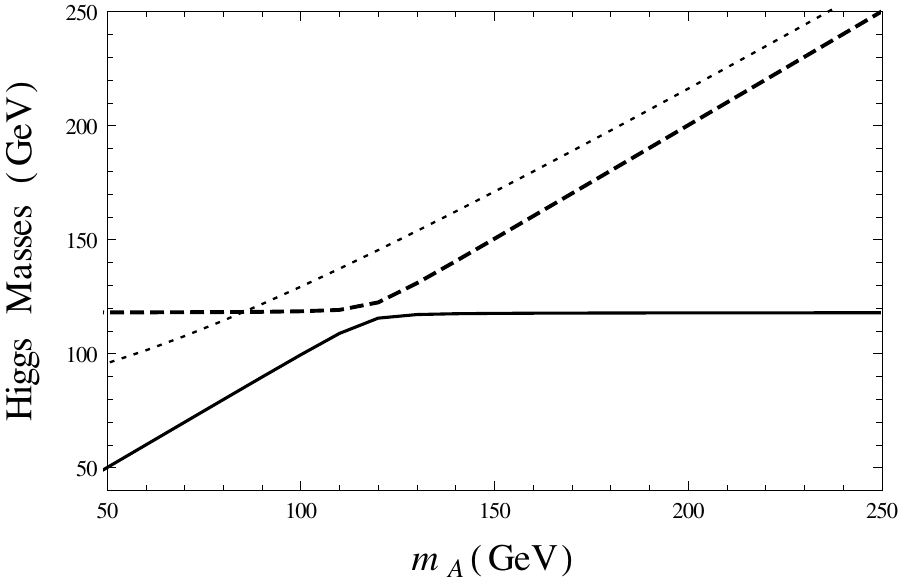}\hspace{2pc}%
\caption{\small{One-loop predictions of the Higgs boson masses $m_h$ (solid), $m_H$ (dashed) and $m_{\pm}$~(dotted) as a function of the CP-odd Higgs boson mass $m_A$. We have considered $tan\beta = 30$ in a scenario where $M_{SUSY}=1~TeV$ and $\mu=200~GeV$.}}
\label{fig:PlotMhmA1L}
\end{figure}
The state of art of the quantum corrections to the lightest Higgs boson mass in the MSSM is really advanced and widely studied. The reported results include fixed-order diagrammatic calculations, resummation of the large logarithms in effective field theories (EFT) and hybrid fixed-order and EFT calculations up to three-loop level. All of these computations have been implemented in several public codes that provide a precise numerical evaluation of the Higgs masses in SUSY models. The relevant references and a comprehensive overview of the current status of Higgs-mass calculations in supersymmetry can be found in \cite{Kuts} and references therein. For the rMSSM, the dominant contributions come from the top-stop Yukawa corrections at zero external momentum in the EW gaugeless limit, which are proportional to the fourth power of the top mass, $M_t$, and to the logarithm of the physical stop masses, $\tilde{m}_{t_1}$ and $\tilde{m}_{t_2}$:
\begin{eqnarray}
    & & \delta_t^{(1)} m_h^2 = \dfrac{3G_F}{\sqrt{2}\pi^2s^2_{\beta}}\, M_t^4 \, \mathrm{ln}\left(\dfrac{\tilde{m}_{t_1}\tilde{m}_{t_2}}{M_t^2} \right). \label{eq:D1Mhdom}
\end{eqnarray}
The source of this correction is the soft breaking of supersymmetry which produces an incomplete cancellation between virtual top and stop loops. In this approximation the one-loop prediction of the Higgs boson masses can be expressed in the simple compact form: 
\begin{eqnarray}
 &  & m_{h}^{2}=\frac{1}{2}\left[m_{A}^{2}+\delta^{(1)}_{t}m_{h}^{2}-\sqrt{\left(m_{A}^{2}+\delta^{(1)}_{t}m_{h}^{2}\right)^{2}-4m_{A}^{2}s_{\beta}^{2}\delta^{(1)}_{t}m_{h}^{2}}\right], \nonumber \\
 &  & m_{H}^{2}=m_{A}^{2}-m_{h}^{2}+\delta^{(1)}_{t}m_{h}^{2}\quad;\quad m_{\pm}^{2}=m_{A}^{2}+M_{W}^{2}. \label{eq:1LmhH}
\end{eqnarray}
The Higgs mass can be also affected by a potentially large stop mixing $X_t$ due to the non-leading effects of the one-loop correction:
\begin{eqnarray*}
 &  & \delta^{(1)}_{X_{t}}m_{h}^2=\frac{3G_{F}M_{t}^{4}X_{t}^{2}}{2\sqrt{2}\pi^{2}s_{\beta}^{2}}\left(2f\left(\tilde{m}_{t_{1}}^{2},\tilde{m}_{t_{2}}^{2}\right)+\frac{2-\left(\tilde{m}_{t_{1}}^{2}+\tilde{m}_{t_{2}}^{2}\right)f\left(\tilde{m}_{t_{1}}^{2},\tilde{m}_{t_{2}}^{2}\right)}{\left(\tilde{m}_{t_{1}}^{2}-\tilde{m}_{t_{2}}^{2}\right)^{2}}X_{t}^{2} \right), \label{eq:D1MhXt}
\end{eqnarray*} 
where the function $f$ is defined as:
\begin{eqnarray}
 & & f\left( x, y\right) = \dfrac{1}{x-y}\mathrm{ln}\left(\dfrac{x}{y}\right). \label{eq:fxydef}
\end{eqnarray}
The dependence of the CP-even and charged Higgs boson masses on the parameter $m_A$ in eq. \ref{eq:1LmhH}, including the dominant one-loop radiative corrections in a scenario where the squarks masses are put at the same supersymmetric scale $M_{SUSY}$, produces the curves in Figure~\ref{fig:PlotMhmA1L}. Note that, independently of $tan\beta$, the masses of the heavy Higgs particles, $m_H$ and $m_{\pm}$, grow linearly without boundary when the scale $m_A$ grows and have approximately the same magnitude, $m_H \approx m_{\pm}$. The lightest Higgs boson mass instead approaches to an asymptotic value, showing a more regular dependence on $m_A$ for large $tan\beta$, where $m_h \simeq m_A$ for small $m_A$ while $m_h \simeq const $ for large $A$-mass. The non-leading effects of the stop mixing $X_t$ allow a simple determination of an upper bound on the lightest Higgs boson mass at one-loop level, yielding the expression:
\begin{eqnarray}
& & m_{h}^{2}\leq M_{Z^0}^{2}c_{2\beta}^{2}+\delta_{t}^{(1)}m_{h}^{2}s_{\beta}^{2}+\delta_{X_{t}}^{(1)}m_{h}^{2}s_{\beta}^{2}. \label{eq:1lMhbound}
\end{eqnarray}
If one considers large values of the parameter $tan\beta$ ($tan\beta \gg 30$) in the scenario of maximal stop mixing, where the value of $X_t$ makes $m_h$ maximal (frequently referred in literature as the $m_h^{max}$ scenario), a general upper bound given by $m_h \lesssim 140~GeV$ is found out. Fortunately for the MSSM, the Higgs boson was discovered at the LHC within this energy region. The contributions of eq. (\ref{eq:D1Mhdom}) and eq. (\ref{eq:D1MhXt}) contain the bulk of the one-loop corrections. The subdominant contributions coming from higher-loop corrections can essentially be reduced to higher-order SQCD effects. 
\begin{figure}
\centering
\subfloat{
\includegraphics[width=14pc]{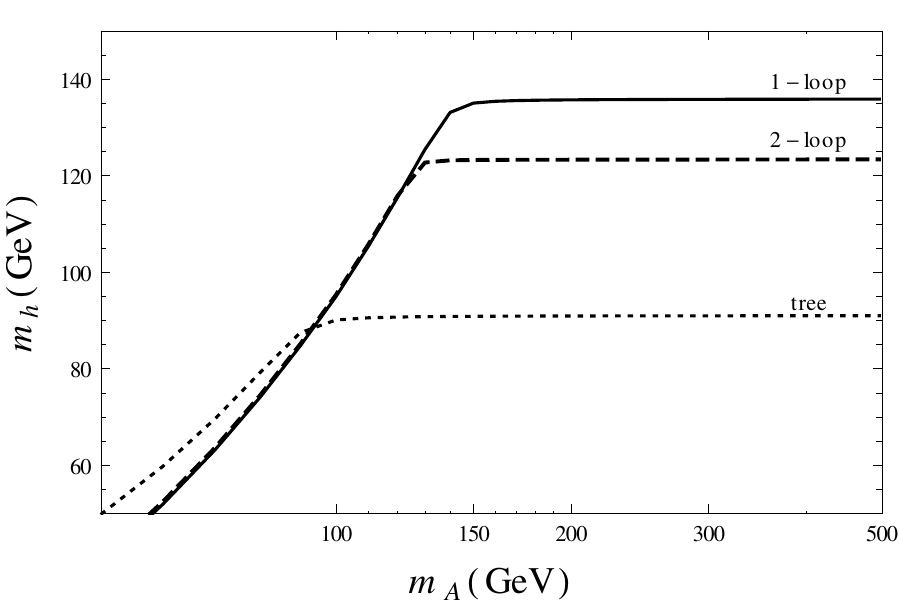}}
\subfloat{
\includegraphics[width=14pc]{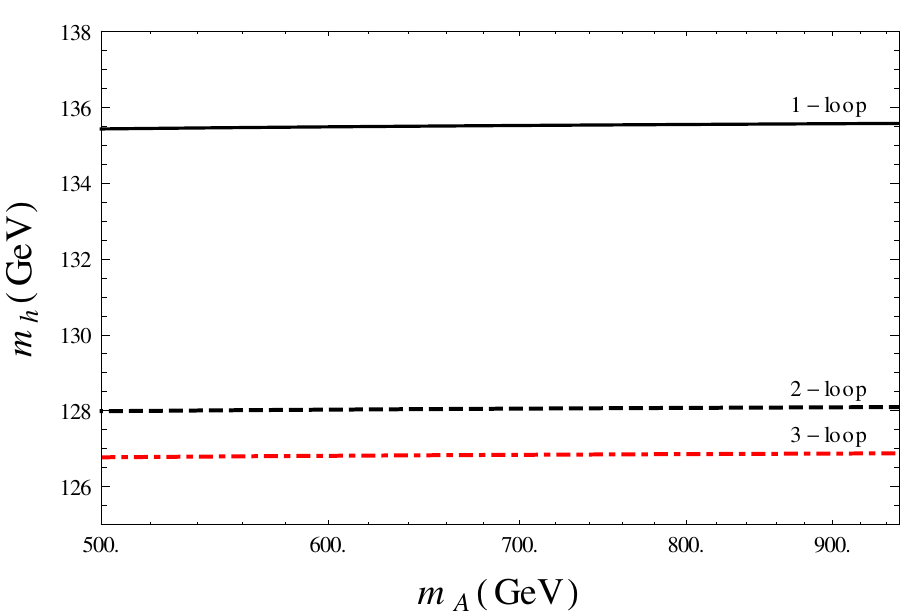}}
\caption{\small{\textbf{Left}: Predictions of the lightest Higgs boson mass, $m_h$, at tree (dotted), one-loop (solid) and two-loop (dashed) level as a function of the CP-odd Higgs boson mass~$m_A$. We have considered the $m_h^{max}$ scenario ($X_t \approx 2.4M_{SUSY}$) at the kinematical point $tan\beta = 30$, $M_{SUSY}=1~TeV$, $\mu=200~GeV$ and with a gluino mass $M_{\tilde{g}}=1.5~TeV$. \textbf{Right}: The three-loop $m_h$-predictions (dot-dashed red line) are included for large $m_A$ in the maximal stop mixing scenario.}}  
\label{fig:PlotMhmA3L}
\end{figure}
Figure \ref{fig:PlotMhmA3L} - \textbf{Left} shows the dependence of the tree-level, full one-loop and leading QCD two-loop $m_h$-predictions coming from \textit{FeynHiggs}~\cite{FeynHiggs} as a function of the scale $m_A$ in the maximal stop mixing scenario for the same election of parameters as in Figure~\ref{fig:PlotMhmA1L}. In this plot one can clearly see the good behaviour of the perturbative expansion. If one looks at the numerical difference between the dashed and the dot-dashed curve as well as the difference between the dashed and the solid line for large $m_A$, one can easily conclude that the size of the one-loop corrections (estimated to be about $45~GeV$) are higher than the size of the two-loop ones (roughly $12~GeV$), as expected from a perturbative quantum field theory. Besides, the two-loop corrections give negative contributions in contrast to the one-loop corrections which are positive. In the right panel is depicted the three-loop predictions of $m_h$ at order $\alpha_t\alpha_s^2$~\cite{Edilson1, Edilson2} as a function of $m_A$ (dot-dashed red line) in the $m_h^{max}$~scenario. In general, the complete $m_h$-prediction is built up as the sum of different contributions. The tree-level value, which accounts for about 60\% of the renormalized mass, the one-loop correction, which represents about 32\% of $m_h$, the two-loop correction, contributing with about 6\%, and finally the three-loop contribution representing about 1\% of the total mass. The size of the three-loop quantum corrections are of the order of $1~GeV$ and have an opposite sign regarding the one-loop corrections for small values of $M_{SUSY}$ of about~$1~TeV$. However, due to the lack of experimental evidence for SUSY particles at this energy, we have to consider benchmark scenarios with even higher SUSY energy scales. Considering the limit where all the soft SUSY-breaking masses as well as the CP-odd Higgs mass ($m_A$) lie around the characteristic scale $M_{SUSY}$, the dependence of the $m_h$-predictions on this scale can be studied.
\begin{figure}
\centering
\subfloat{
\includegraphics[width=14pc, height=10pc]{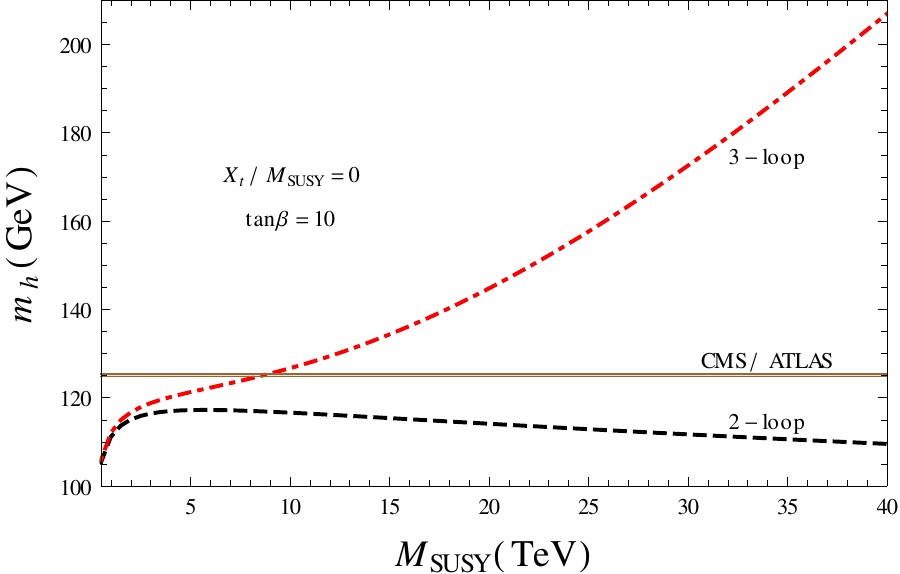}}\hspace{10pt}
\subfloat{
\includegraphics[width=12pc, height=11pc]{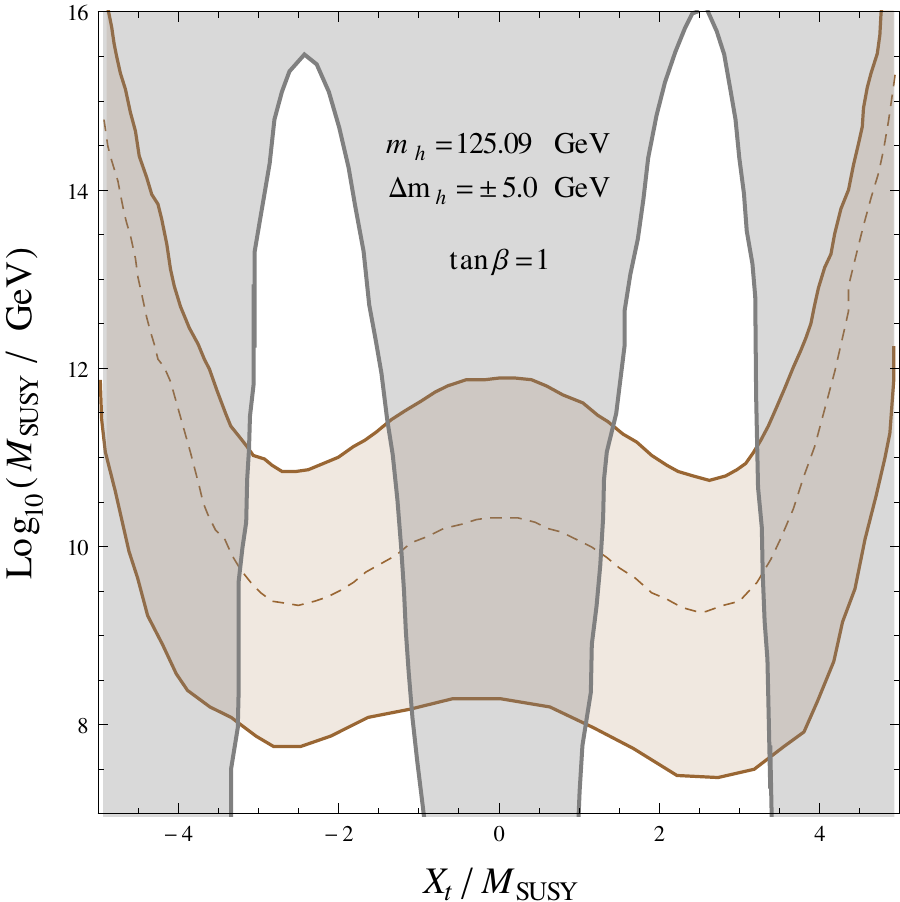}}
\caption{\small{\textbf{Left}: Dependence of $m_h$ on the supersymmetric scale $M_{SUSY}$ for
a vanishing stop mixing ($X_t = 0$) at full two-loop (black dashed line) and O($\alpha_t \alpha_s^2$) three-loop level (red dot-dashed line). The brown band is the CMS/ATLAS Higgs boson
mass: $125.09 \pm 0.24 \, GeV$. \textbf{Right}: Higgs boson mass predictions as a function of $M_{SUSY}$ and $X_t$. The brown region corresponds to the points compatible with a Higgs boson mass of $125.09 \pm 5.0 \, GeV$ for $tan\beta = 1$. The gray region represents the values of $X_t$ and $M_{SUSY}$ where $\lambda(M_{SUSY})$ becomes negative.}}  
\label{fig:NewPlots}
\end{figure}
In Figure~\ref{fig:NewPlots} - \textbf{Left} we present a numerical comparison of our three-loop fixed-order predictions of $m_h$ to the two-loop predictions coming from FeynHiggs for large SUSY scales in the range $0.5\, TeV\leq M_{SUSY}\leq 40\, TeV$. The dependence of $m_h$ on $M_{SUSY}$ is evaluated for a vanishing stop mixing, $X_t=0$, and for $tan\beta = 10$. The two and three-loop fixed-order results are represented with the black dashed and the red dot-dashed lines respectively. The brown band is the experimental Higgs boson mass measured at LHC and its corresponding uncertainty. Note that for scales above $10\, TeV$ the difference between the two-loop and three-loop results rapidly increases up to about $25 \, GeV$ when $M_{SUSY}$ grows to up to $20\, TeV$ and grows monotonically reaching about $100\, GeV$ at $M_{SUSY} = 40\, TeV$. This pronounced behaviour depends crucially on our election of the input parameters and is a consequence of the presence of $n$-loop logarithms of the form $log^n (M_{SUSY}/M_t)$ in the three-loop fixed-order Higgs self-energy corrections. Those $n$-loop logarithms are known as large logarithms because they are the source of additional large contributions in the three-loop predictions of $m_h$ when there is a large hierarchy between the EW scale $M_t$ and the SUSY scale $M_{SUSY}$. The large logarithms contributions can spoil the convergence of the perturbative expansion and yield unreliable predictions of the Higgs boson mass. A fixed-order computation is thus recommended for low values of $M_{SUSY}$ not separated too much from $M_t$. For high SUSY energy scales, a resummation of the large logarithms with the help of the renormalization group techniques is required as was discussed in~\cite{Edilson3}. Once the resummation is performed, the dependence on $M_{SUSY}$ can be evaluated up to super-high energy scales of the order of the GUT (SUSY Grand Unified Theories) energy. In Figure~\ref{fig:NewPlots} - \textbf{Right} we further explore the dependence of the Higgs boson mass on the SUSY scale and the stop mixing parameter. We have plotted the region of parameters which is compatible with a Higgs boson mass equal to $125.09 \pm 5.0 \, GeV$ (brown region) for large values of $M_{SUSY}$ between $10^7 \, GeV \lesssim M_{SUSY}\lesssim 10^{16}\, GeV$ and for $tan\beta = 1$. As was argued also in~\cite{Edilson3}, it is possible to fix upper bounds on the SUSY scale $M_{SUSY}$ for scenarios with large values of $tan\beta$ ($\tan\beta > 10$) if we impose that the $m_h$-predictions in the MSSM is compatible with the combined CMS/ATLAS measured Higgs boson mass within the actual combined uncertainties. However, for small values of $tan\beta$ ($tan\beta < 10$) such upper bounds coming from the LHC measurements cannot be derived and a strong dependence between $M_{SUSY}$ and $X_t$ is observed. In fact, for large values of the stop mixing parameter, $| X_t | \gtrsim 3M_{SUSY}$, the $m_h$-prediction is compatible with the LHC measurement for $M_{SUSY} \gtrsim 10^7 \, GeV$. It is possible, however, to derive upper bounds over $M_{SUSY}$ for small $tan\beta$ values if we impose the positivity of the running Higgs self-coupling $\lambda(Q)$ from the EW scale until the next threshold, $Q=M_{SUSY}$, which is equivalent to impose the stability of the SM renormalization group improved Higgs effective potential, 
\begin{equation}
    V(h)^{RGI} \approx \frac{\lambda(Q)}{4}h^4(Q), \label{eq:RGI-EP} 
\end{equation} 
up to the scale where supersymmetry is supposed to appear. The SM is considered here as a low-energy EFT of the rMSSM valid in the limit where we integrate out all SUSY QCD particles at the scale $M_{SUSY}$, and therefore the RGEs correspond to the three-loop SM beta functions while the threshold corrections to the boundary condition for the Higgs quartic coupling contain two-loop corrections including the MSSM particles. The full two-loop threshold corrections to $\lambda$ can be found in~\cite{Bagnaschi} and references there in. In the gray region of Figure~\ref{fig:NewPlots} - \textbf{Right} we have pictured the points where the quartic Higgs self-coupling is negative at the SUSY scale, $\lambda(M_{SUSY}) < 0$, leading to an unstable electroweak vacuum. If the stability of the EW vacuum is required, then the parameter space in the gray region are excluded and the intersection of the gray border line with the brown region could give constraints on the $M_{SUSY}$ value as a function of $m_h$. Note that the region of exclusion occurs in general around $|X_t/M_{SUSY}| = 0$ and for large values of $X_t$, $|X_t/M_{SUSY}| > 3$. Specifically, for a $125.09\, GeV$ Higgs boson mass (dashed brown line) and a stop mixing of $|X_t/M_{SUSY}| = 1.2$, the intersection of the gray curve with the dashed brown line provides the upper bound $M_{SUSY} \lesssim 10^{10}\, GeV$. Assuming an uncertainty of $\pm 5\, GeV$ on the Higgs boson mass, this bound can reach a maximum value of $M_{SUSY} \lesssim 10^{11} \, GeV$ for $|X_t/M_{SUSY}| \approx 1.4$ as was also verified in~\cite{Edilson2, Allanach}. However, such a region of superhigh SUSY energy scales cannot be tested at the current LHC or dark matter underground experiments and at the future colliders. An experimental test requires, therefore, the design of experiments capable to provide indirect signals of SUSY particles belonging to models for high energies of the order of the GUT scale, as is the case of supergravity. This has brought a special interest of the study of supersymmetry in the context of astroparticle physics and cosmology. The cosmological consequences and the possible applications to SUSY models can be found in~\cite{cosmo1,cosmo2,cosmo3,cosmo4,cosmo5}. It is worth mentioning that this super-high energies are not the only target where we could expect some SUSY evidence. There are recent phenomenological analysis that show the possibility to reproduce the $4.2\sigma$ discrepancy in the muon (g-2) anomaly together with the full Dark Matter (DM) relic density of the universe by taking the lightest supersymmetric particle (LSP) to be the DM candidate~\cite{Chakra}. An upper limit of $\sim 600 \, GeV$ is obtained in this work for the LSP and next-to (N)LSP masses establishing clear search targets for the future HL-LHC EW searches and for future high-energy $e^{+}e^{-}$ colliders. This bound is in accordance with the LHC exclusion bounds found for searches of direct production of charginos, neutralinos and sleptons~\cite{ATLAS1,ATLAS2,ATLAS3,ATLAS4}. We would like to remark that for the phenomenological MSSM (pMSSM) parameter ranges, where the SUSY masses are of the order of 1 to 10 TeV, the~Higgs coupling measurements with the accuracies obtained on the LHC Run 2 data and those expected for the HL-LHC and future $e^{+}e^{-}$ colliders can exclude up to 20\% of the accepted pMSSM points, as~was recently shown in~\cite{Djouadi}. Therefore, a~large part of the pMSSM parameters are still to be probed, suggesting the conclusion that the properties of the observed Higgs boson are also MSSM-like. Finally, we stress that there are also well-motivated non-supersymmetric theories that naturally accommodate the experimental observation of a light 125 GeV Higgs boson and exhibit a large separation of scales able to match experimental observations, as~is the case with the composite Higgs models~\cite{Redi, Bellazzini, Dobado} or the low-energy effective Higgs theories, such as~the two Higgs doublet model (2HDM)~\cite{Branco, Bernon}.  

\section{Theoretical Uncertainties and Next to Minimal Supersymmetry} \label{HiggsNMSSM}

\begin{figure}
\centering
\includegraphics[width=25pc]{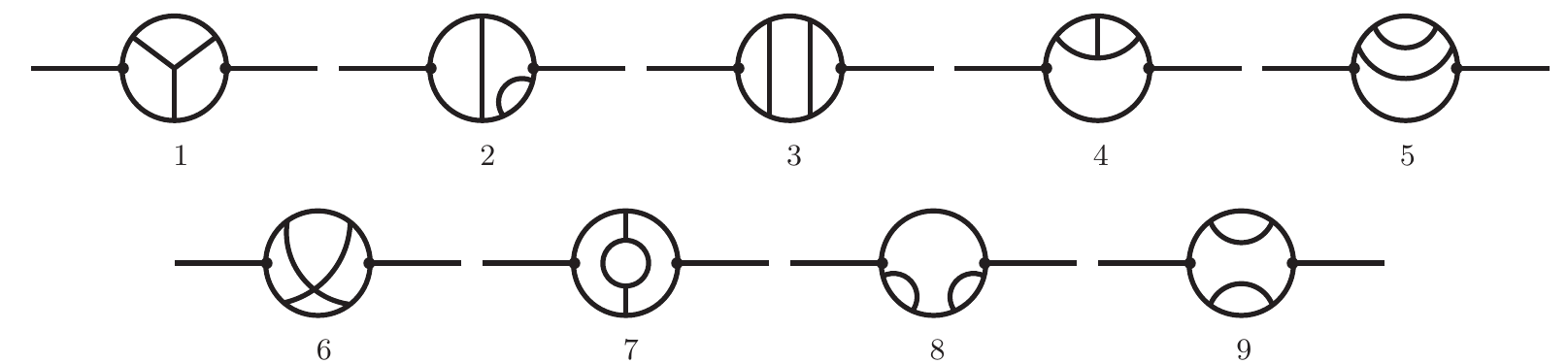}\hspace{2pc}%
\caption{\small{Self-energy topologies contributing to the three-loop corrections to the Higgs boson mass at order O($y_t^6$) in the SM. The topology 6 is the only non-planar self-energy.}}
\label{fig:Self-energies}
\end{figure}

The computation of the Higgs boson mass in SUSY models requires to impose several approximations, including, in particular, the truncation of the involved perturbative expansions at some loop level, the limit where the effects of some subset of couplings within the considered loop-order are neglected or the approximation where some of the kinematics invariants vanish, as in the case of the effective potential approach where the limit of vanishing external momentum is adopted. Therefore, an estimation of the theoretical uncertainty, which measures the possible effects of the missing terms, must be included in a realistic computation of $m_h$. For the case of the MSSM, the Higgs boson mass has been computed at fixed-order by using a Feynman diagrammatic approach, but also with the help of the renormalization group equations in an effective field theory (EFT) approach. Due to the well-known large logarithmic effects in the fixed-order calculation, an EFT computation where the large logarithms are resummed is considered the best option for high SUSY scales larger than about $1~TeV$~\cite{Edilson3}. In fact, pure fixed-order calculations are more reliable when the SUSY masses are close to the EW scale, while pure EFT calculations are more reliable in heavy-SUSY scenarios. As a consequence, hybrid calculations where the fixed-order and the EFT results are combined in order to improve the theory uncertainty at each kinematical point have been implemented by different groups. Currently there are public codes which allows a numerical estimation of the different theoretical errors associated to the EFT, fixed-order and hybrid calculations of the Higgs boson mass. For an extensive and detailed explanation of these estimations, see~\cite{Kuts} and references therein. Nonetheless, in the following we highlight some of the uncertainty sources which will require in the future improved perturbative computations. \\ In the diagrammatic approach, a full two-loop fixed-order calculation of $m_h$, of the lightest Higgs boson  is still missing. The two-loop computation in the SQCD sector is complete, but the effects due to the EW gauge couplings are only available in the limit of vanishing external momentum. A complete two-loop calculation of scalar tadpoles and self-energies of a general renormalizable theory was presented recently in~\cite{Goodsell}; however, the relations between running parameters and on-shell observables require also the full two-loop gauge boson self-energies which are still missing. At three-loop level, contributions exist in the effective potential approach at order O($\alpha_t\alpha_s^2$) while the calculation of the three-loop corrections that involve lower powers of the strong gauge coupling, e.g. O($\alpha_t^2\alpha_s$), O($\alpha_t^3$), contains only the logarithmic effects. On the other hand, in the EFT calculation of $m_h$, beyond next to leading logarithmic (NLL) order, the estimation has so far been performed only for the simplest heavy-SUSY scenario where the MSSM is matched directly to the SM and all the SUSY particles are putted at the same scale $M_{SUSY}$ in the limit $M_{SUSY}\gg M_t$. The next to next leading logarithms (NNLL) corrections neglect most of the effects that involve the EW gauge couplings, while at NNNLL they account only for the effects that involve the top Yukawa coupling combined with the highest powers of the strong gauge coupling. Taking into consideration these sources of uncertainties together with the additional theoretical and parametric uncertainties described in references~\cite{Kuts, FeynHiggs2, FlexibleEFTHiggs1, FlexibleEFTHiggs2}, the codes \textit{FeynHiggs} and \textit{FlexibleEFTHiggs} allow for an estimation of the Higgs boson mass uncertainty as a function of the SUSY scale. 
\begin{figure}
\centering
\includegraphics[width=16pc]{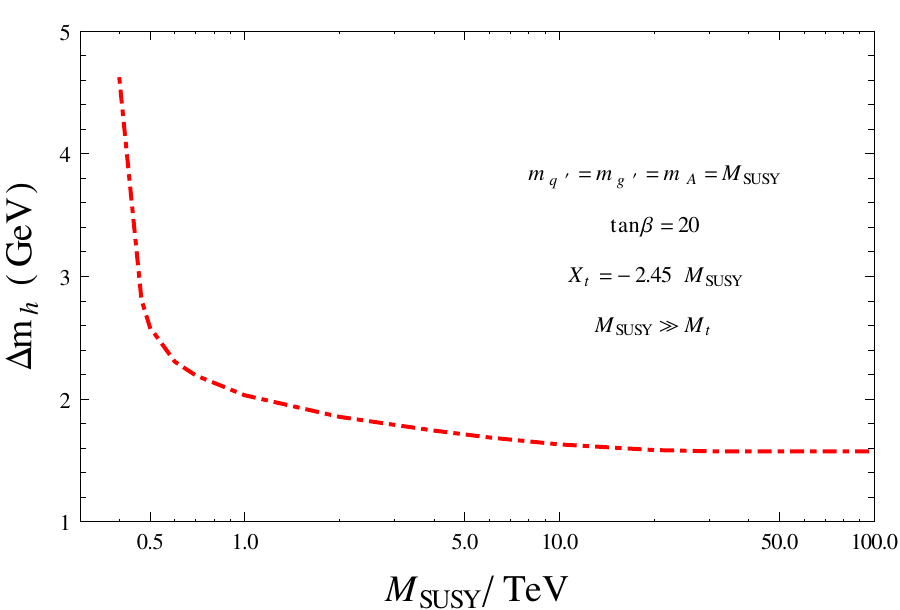}\hspace{2pc}%
\caption{\small{Estimation of the theory uncertainty of the Higgs mass produced with \textit{FeynHiggs} in the hybrid approach for a MSSM scenario with degenerate SUSY masses.}}
\label{fig:PlotUncertainty}
\end{figure}
In Figure~\ref{fig:PlotUncertainty} an estimation of the theory uncertainties in the hybrid approach of \textit{FeynHiggs} as a function of the SUSY scale is drawn. The calculation was done in a scenario where all the squark masses ($m_{q'}$), the gluino mass ($m_{g'}$), the parameter $\mu$ and the CP-odd Higgs boson mass $m_A$ are set equal to a common scale $M_{SUSY}$ which is varied from $0.4~TeV$ to $100~TeV$. The stop mixing parameter is taken to be $X_t=-2.45M_{SUSY}$ while $tan\beta=20$.  Besides, the squark masses and $X_t$ are running parameters evaluated at the scale $Q=M_{SUSY}$ in the $\overline{DR}$ renormalization scheme~\cite{DRbar}. Assuming this MSSM scenario, the total theoretical uncertainty is in the range $\Delta m_h\approx 1-5~GeV$, where, in particular, for large values of $M_{SUSY}$ the SUSY scale dependence is mild and the uncertainty decreases up to about $1.5~GeV$. The size of $\Delta m_h$ is significantly large compared with the current experimental error estimated at LHC ($\sim 200~MeV$) which is about an order of magnitude less than the theoretical uncertainties estimated in Figure~\ref{fig:PlotUncertainty}. Therefore, the theory uncertainty must be reduced by about a factor of $10$ in order to reach the current and future experimental precision. This aim can be achieved by improving the accuracy in the measurement of the EW observables at the future colliders, but also through the explicit computations of the dominant missing terms in the theoretical prediction of $m_h$, which additionally could tell us whether the current estimation of $\Delta m_h$ is too optimistic or too pessimistic. If the accuracy of the Higgs mass calculation is improved, the existing uncertainty estimate must be adapted so that it simulates the dominant terms among those which have not been yet computed. Any future improvement in the accuracy of the fixed-order and EFT calculations of $m_h$ would also eventually improve the uncertainty in the hybrid calculations. \\ There are several strategies to make an improvement. We just refer to one of them in this article which is related to a work in progress that our group is currently doing. The three-loop computation of the Higgs mass, in the SM as well as in the MSSM, was done in the effective potential approach which is equivalent to compute the self-energy corrections in the limit of vanishing external momentum. However, the inclusion of the external-momentum effects in the three-loop self-energies is actually attainable by numerical methods, based on the dispersion relations, implemented in \textit{TVID2}~\cite{TVID2}, that allows the evaluation of three-loop planar self-energies with arbitrary masses and external momentum. In addition, with the results presented recently in~\cite{Martin3}, where the differential equations method is used to compute numerically the renormalized $\epsilon$-finite master integrals for arbitrary external momentum invariant and, in principle, for arbitrary masses, the evaluation of non-planar self-energies is also possible. For a perturbative computation at order O($y_t^6$) in the SM, which is the non-QCD dominant contribution, planar as well as non-planar self-energies with just cubic vertices are required, as you can see in Figure~\ref{fig:Self-energies}. 
The external lines of the topologies contain only the Higgs field ($h$). The internal lines (propagators) contain the top quark ($t$), bottom quark ($b$), Higgs and Goldstone bosons ($G^{0}$ and $G^{\pm}$) fields. The cubic vertices involved in the computation are $hht$, $G^{0}G^{0}t$ and $G^{\pm}tb$. In the Landau gauge, therefore, the Feynman integrals contain four different scales, namely, the Higgs boson mass, the top quark mass, the bottom quark mass and the external momentum of the self-energies. The mass of the bottom quark is considered here as a small variable which is maintained as a non-zero input parameter inside the propagators of the amplitudes, the latter with the aim of avoiding additional infrared divergences, but its contribution to the vertex $G^{\pm}tb$ is disregarded when appears in the numerators of the integrals. In particular, Feynman diagrams involving the charge Goldstone boson $G^{\pm}$ contain Dirac traces with the chiral matrix $\gamma_{5}$ and an arbitrary number of gamma matrices. Due to the presence of four independent momentum scales, the external momentum and the three loop momenta, the self-energy diagrams can contain contributions of Dirac traces with a single $\gamma_{5}$ and four $\gamma$ matrices. The details of the regularization of these amplitudes and the consequent reduction and explicit evaluation of the resulting master integrals coming from the topologies in Figure~\ref{fig:Self-energies} will be presented in a future publication. An extension of this calculation to the MSSM in a suitable regularization scheme that preserves supersymmetry would require also diagrams with quartic vertices involving the stop fields and will allow an evaluation of the non-logarithmic terms at order O($y_t^6$). In fact, even if in the MSSM supersymmetry is explicitly broken, the mechanism is soft therefore the supersymmetric cancellation are more efficiently realized in a regularization scheme, preserving supersymmetry \cite{Tod}.  The inclusion of the external-momentum effects in the self energies is relevant when going beyond the gaugeless limit in the MSSM, or in models with additional Higgs self-couplings. The estimations of those effects in the SQCD sector has to be still determined. 

Although the MSSM is the BSM model which has the most precise theoretical predictions of the Higgs boson mass and their associated uncertainties, there are phenomenological indications that suggest the necessity of an extension of its Higgs sector. The searches of BSM Higgs bosons at the RUN2 of the LHC have shown several excesses in the data, from which two facts are remarkable. The first one is the excess showed in several events above the background expectation around the same mass scale of a hypothetical new Higgs boson with a mass of $400 GeV$~\cite{excess1,excess2,excess3,excess4}. The second one, take into account the possibility of a BSM Higgs boson with a mass lighter than the $125~GeV$ scale, due to the local excess observed around the scale of about $96~GeV$ by the CMS~\cite{CMSexcess}, confirmed by the LEP experiments~\cite{LEPexcess} as well. In previous analyses focussing just on these two excesses it was shown that in the MSSM neither a $400~GeV$ Higgs boson nor the CMS excess at around $96~GeV$ can be realized~\cite{Bechtle}. Extended SUSY models instead, like the NMSSM~\cite{NMSSM1,NMSSM2} or the $\mu\nu$SSM~\cite{ref-munuSSM1,ref-munuSSM2}, can account for the excesses at a level of roughly $1\sigma$. An analysis of the possible interpretations of the observed excesses in the context of the N2HDM and the NMSSM models can be found in~\cite{Thomas}. A complete explanation of the excesses found by LEP-II and CMS at 96 GeV in the natural NMSSM (nNMSSM) can be also found in~\cite{Junjie1, Junjie2}. Additionally, the minimal SUSY extension of the SM cannot explain the neutrino oscillations and is compatible with the $4.2\sigma$ anomaly in the muon $g-2$ experiment just for a compressed region of EW parameters. The anomaly could be explained in terms of light EW SUSY particles, as the sleptons or the neutralino, with masses between $\sim 250~GeV$ and $\sim 700~GeV$. If those BSM particles are discovered in this range, then the motivation for improving the theoretical accuracy of the MSSM Higgs-mass prediction will be even stronger. Otherwise, a next to MSSM model suitable to explain the muon $g-2$ anomaly must be considered. This issue has been anticipated in recent works. There is an interesting proposal where the MSSM Higgs sector is extended in order to have four Higgs doublets, two of which couple just to quarks and leptons of the third generation, whereas the other two have much smaller vacuum expectation values and provide masses to the first and second generation. The model is called FSSM~\cite{FSSM1, FSSM2} and its region of parameters includes sleptons in the multi-TeV regime, beyond the current reach of the LHC, capable of solving the muon $g-2$ anomaly. In particular, the extended Higgs sector of the FSSM can account the experimental $(g-2)_\mu$ results in scenarios where the stops need to be considerably heavier than smuons, higgsinos and EW gauginos. Another interesting alternative is the $\mu\nu$SSM~ which is a highly predictive model that can explain the neutrino oscillations via an EW seesaw mechanism~\cite{ref-munuSSM3}, provide a solution of the so-called $\mu$-problem~\cite{ref-munuSSM4} and solve the muon $g-2$ anomaly~\cite{ref-munuSSM5} simultaneously. Recently, a vacuum stability analysis that takes into consideration the possibility of long-lived metastable vacua of the neutral scalar potential in the $\mu\nu$SSM was also presented~\cite{ref-munuSSM6}. The matter content of this model is enlarged to include right-handed neutrinos and their corresponding superpartners, the “right-handed” sneutrinos, that are gauge singlet scalar fields. \\ Finally, we point out that, in order to accommodate the observed phenomenology, the MSSM parameters must be enlarged with additional degrees of freedom. This can go against the aim to improve the theoretical uncertainty of the Higgs boson mass since the next to MSSM models contains more particles and additional large couplings. Thus, an estimation of the theoretical uncertainties are also relevant in this context. The application of the existing techniques for the MSSM Higgs mass uncertainty calculations should be useful to accomplish this aim.  

\section{Conclusions} \label{Conclusions}

In this article, we have presented a general review of the precision calculations of the Higgs boson mass in the SM ($M_h$) as well as in the rMSSM ($m_h$). The tree-level expressions of the SM and MSSM Higgs potential together with a discussion about the effects of the dominant quantum corrections have been also reviewed. The theoretical uncertainty associated to $M_h$ in the SM can be as large as $1~GeV$ while for $m_h$ the uncertainty estimated in \textit{FeynHiggs} and confirmed in \textit{FlexibleEFTHiggs} is of the order of $1-5~GeV$. In both cases, an improvement of the precision in the computation of the Higgs boson mass is required since the current experimental uncertainty obtained at LHC is $\sim 0.2~GeV$ and is expected to reach a value of about $50~MeV$ in future colliders. We have described the main sources of theoretical uncertainties in both models. In particular, we have focused our attention in the three-loop corrections to the Higgs boson mass where the fixed-order calculations based on the Feynman diagrammatic approach have been reported in the limit of zero external-momentum for the Higgs self-energies. We have proposed a way to improve the accuracy of the calculation based on the possibility to estimate the external-momentum effects that is reachable by the recent results presented in~\cite{TVID2, Martin3}. On the other hand, the SM and the MSSM predictions are moving away from the current observed phenomenology. We have discussed briefly the alternative models which can accommodate the recent results obtained at the RUN2 of the LHC and the muon $g-2$ anomaly found at FERMILAB. In particular, these alternatives include some extensions of the minimal version of the supersymmetric standard model which enlarge the Higgs sector with additional large parameters. This could go against the intention to improve the theoretical uncertainty on the Higgs boson mass and therefore, as a research perspective, we consider relevant to estimate the theory uncertainties for the next to MSSM models discussed in Section~\ref{HiggsNMSSM}. We hope that this review will stimulate the scientific discussion  with groups interested in this research line.

\end{document}